# Voltage tunable quantum dot array by patterned Ge-nanowire based metal-oxide-semiconductor (MOS) devices

Subhrajit Sikdar[1], Basudev Nag Chowdhury[1], Rajib Saha[1], Sanatan Chattopadhyay[1,2,*]

[1]Department of Electronic Science, University of Calcutta, Kolkata, India

[2]Centre for Research in Nanoscience and Nanotechnology (CRNN), University of Calcutta, Kolkata, India

Corresponding author Email: *scelc@caluniv.ac.in

**Semiconductor quantum dots (QDs) are being regarded as the primary unit for a wide range of advanced and emerging technologies including electronics [1], optoelectronics [2], photovoltaics [3] and biosensing applications [4] as well as the domain of q-bits based quantum information processing [5]. Such QDs are suitable for several novel device applications for their unique property of confining carriers 3-dimensionally creating discrete quantum states. However, the realization of such QDs in practice exhibits serious challenge regarding their fabrication in array with desired scalability and repeatability as well as control over the quantum states at room temperature. In this context, the current work reports the fabrication of an array of highly scaled Ge-nanowire (radius ~ 25 nm) based vertical metal-oxide-semiconductor devices that can operate as voltage tunable quantum dots at room temperature. The electrons in such nanowire experience a geometrical confinement in the radial direction, whereas, they can be confined axially by tuning the applied bias in order to manipulate the quantum states [6-7]. Such quantum confinement of electrons has been confirmed from the step-like responses in the room temperature capacitance-voltage (C-V) characteristics at relatively low frequency (200 kHz). Each of such steps has observed to encompass convolution of the quantized states occupying ~6 electronic charges. The details of such carrier confinement are analyzed in the current work by theoretically modeling the device transport properties based on non-equilibrium Green's function (NEGF) formalism.**

The advancement of nanotechnology has provided a systematic route for controlled fabrication of several nanostructures including 2D-nanofilms, 1D-nanowires/nanotubes and 0D-quantum dots by adopting top-down or bottom-up approach as well as by hybrid technology [1, 8-10]. Among such miniaturized structures, a special effort has been made to realize the voltage tunable quantum dots (VTQDs) in order to achieve appropriate control over their quantum states **[11-15]**. However, fabricating patterned arrays of such nanostructures of the physical dimensions that exhibit quantum effects at room temperature is still a challenge. The current decade has witnessed several reports on fabricating nanowire arrays, while their radii being of the order of 50 nm are too large for such quantum confinement to be observed **[16-20]**.

In this context, the current work reports the realization of voltage tunable quantum confinement of electrons at room temperature by fabricating an array of vertical Ge-nanowire metal-oxide-semiconductor (Ge-NWMOS) devices on p-Si-substrate. The patterned array of sites for such nanowire formation is scaled down in the current work by optimizing e-beam lithographic process **[see SI]** in order to obtain the nanowire-radii of the order of Ge excitonic Bohr radius (i.e. 24.3 nm **[21]**). The vertical NWMOS devices are fabricated by employing e-beam evaporation technique to deposit Ge and $SiO_2$ sequentially on p-Si substrate, followed by dc sputtering of Pt over it; and for ground contact, Al is deposited on the back face of substrate by thermal evaporation technique **[see SI]**. It is worthy to mention that Ge is chosen as the semiconductor nanowire material due to its large excitonic Bohr radius and $SiO_2$ is selected as insulator since it is expected to reduce tunneling leakage significantly due to its larger electron effective mass **[7]**. On application of a positive bias at the metal terminal of such NWMOS, the electrons are 3-dimensionally confined into the quantum well created at semiconductor/oxide (Ge-NW/$SiO_2$) junction due to: geometrical quantization in two transverse dimensions (i.e. the radial directions of nanowire), and electrostatic quantization along the nanowire-axis. For a particular NWMOS, the positions of quantum states in energy space significantly depend on the shape of such quantum well and the corresponding electrostatic quantization can be manipulated by varying the applied bias leading to the formation of VTQD **[6-7]**.

The schematic of such NWMOS based VTQD is shown in Fig. 1(a) along with the energy band diagram of the device in unbiased condition as depicted in Fig. 1(b). In the current work, the applied voltage is varied in order to study the entire range of so-called 'accumulation-to-inversion' region of MOS capacitance. Generally, the device to exhibit such transition requires the semiconductor to be of specific doping concentration, i.e., p- or n- type. However, fabrication of ultra-scaled semiconductor nanowires of such particular type with controlled doping is technologically very challenging, since even a single dopant results in a very high doping concentration (e.g. ~$10^{16}$/cc) in such nanostructures.

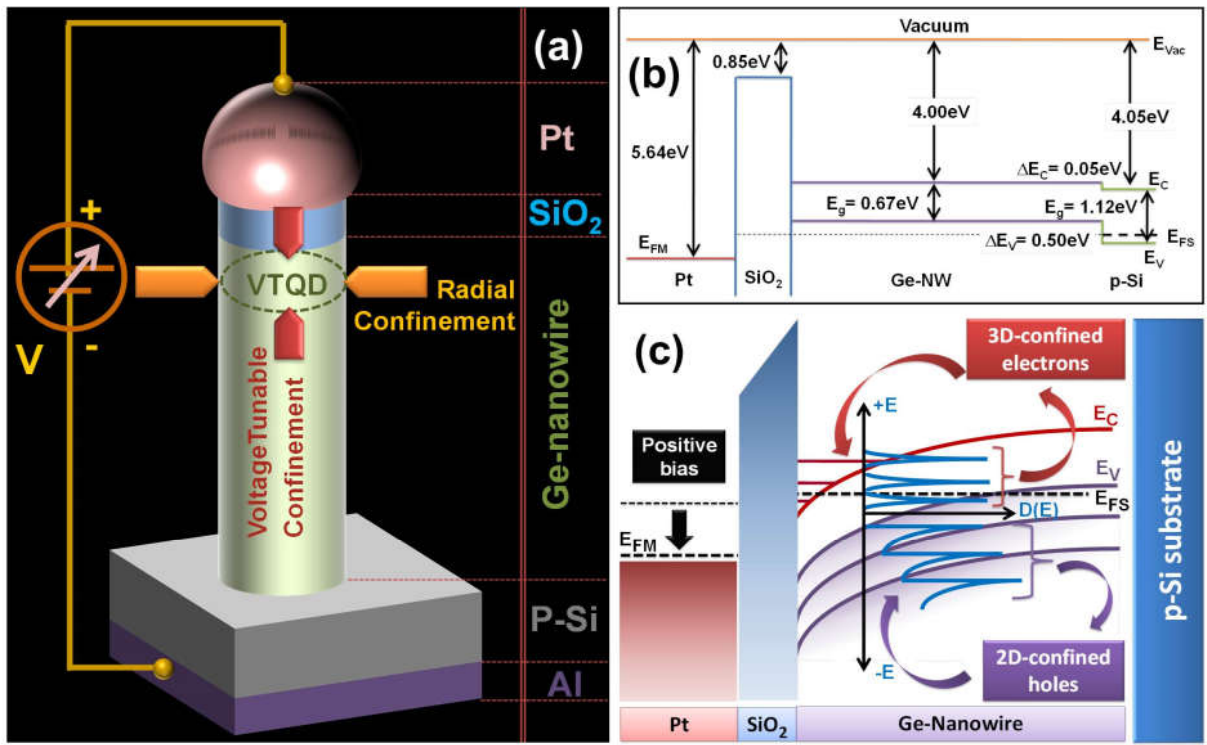

Fig. 1. (a) Schematic of Ge-nanowire based MOS device on Si-substrate indicating the formation of VTQD near Ge-NW/SiO$_2$ interface due to the combined effect of structural and electrical confinement along radial and longitudinal directions, respectively; (b) Energy band diagram of the NWMOS device at unbiased condition illustrating the Si/Ge valence and conduction band offsets; (c) Energy band diagram for positive bias applied on the metal depicting the regions of 3D-confined electrons and 2D-confined holes along with their respective local density of states; the 3D-confined electrons form the VTQD.

In this context, the array of Ge-NWMOS devices is developed in the current work on p-Si substrate based on a scheme of nanowire doping which is conceptually different from the conventional p/n-MOS technology. The intrinsic Ge-nanowire here forms a hetero-junction with p-Si substrate that leads to make it hole-dominated (i.e. p-type) semiconductor, which is attributed to the inherent valence band offset of 0.5 eV at the Ge-NW/p-Si interface as shown in Fig. 1(b).

While applying a negative bias at the metal terminal, such holes remain accumulated within Ge-nanowire leading to develop the 'accumulation' capacitance in the device. However, during the transition from negative to positive bias, such holes tend to move towards the substrate, where the potential barrier at Ge/p-Si interface due to the valence band off-set of 0.5 eV creates a semi-confining region. When further positive bias is applied at the metal terminal of NWMOS, the electrons get confined 3-dimensionally in the quantum well created at semiconductor/oxide (Ge-NW/SiO$_2$) junction due to voltage driven band-bending in the nanowire as illustrated in Fig. 1(c). It is worthy to mention that at such condition (i.e. positive bias), the holes are 2D-confined in radial directions whereas free to move along the nanowire axis, thereby exhibiting the 1-D density of states (DOS); however, the 3D-confined electrons observe the DOS as delta function in energy space around the respective quantum states (see Fig. 1(c)). Therefore, the resulting 'inversion' capacitance is the manifestation of electron confinement in such quantum states created within the NWMOS and can be manipulated by applying appropriate voltage to the device.

The FESEM images of such Pt/SiO$_2$/Ge-nanowire MOS structures fabricated as patterned array on p-Si substrate with various inter-spacing are shown in Fig. 2. The different material regions of such voltage tunable quantum dots are indicated in Fig. 2(a), while Fig. 2(b)-(d) depict the arrays with inter-dot spacing of ~150 nm, ~200 nm and ~250 nm. The average radius of Ge-nanowires is obtained to be ~25 nm, with a standard deviation of ~5 nm [see SI], which corroborates to the excitonic Bohr radius of Ge (24.3 nm). Such variation of inter-dot spacing offers desired array of VTQD based devices as per the requirement of emerging technological applications and large scale production.

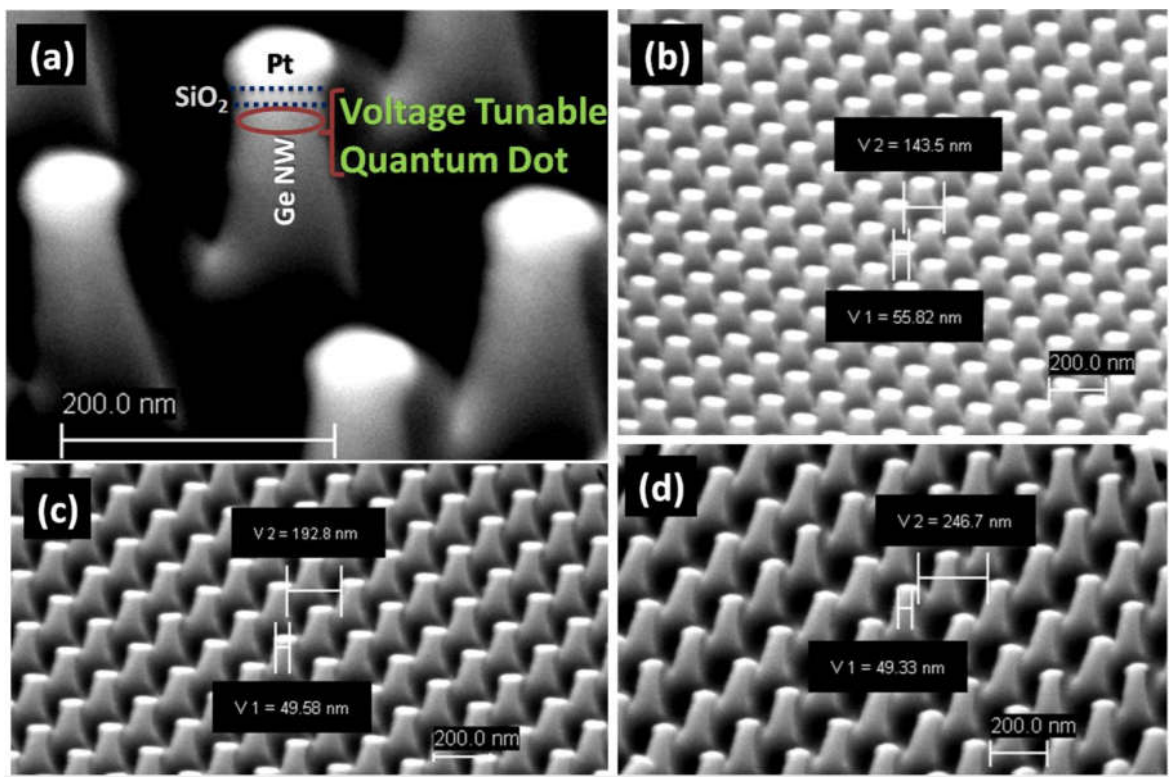

Fig. 2. (a) FESEM image of the Pt/SiO$_2$/Ge-NW vertical MOS device indicating the formation of voltage tunable quantum dot region on top of the Ge-nanowires beneath the oxide/semiconductor interface. The image is captured at a magnification of 120 kX and EHT of 3 kV; FESEM images of the array of NWMOS based VTQD devices with inter-nanowire spacing of (b) ~150 nm, (c) ~200 nm and (d) ~250 nm. Such devices are patterned using electron dose of 65k µc/cm$^2$. The nanowire radii for all the cases are obtained ~25 nm (distribution of the nanowire radii is given in the [SI]).

Fig. 3(a) represents the TEM image of a single NWMOS structure where the different regions of Ge-nanowire, SiO$_2$ and Pt are indicated, whereas the corresponding SAED pattern obtained from the Ge-nanowire is shown in Fig. 3(b). It is apparent from the figure that such SAED spots are arranged in a circular ring which is attributed to the electron diffraction from nanowire [18] and the corresponding d-value indicates [200]-plane of Ge [22]. The elemental content of such single nanostructure is shown in Fig. 3(c) by plotting the characteristic peaks of energy-dispersive X-ray (EDX), measured in-situ under TEM, which confirms the presence of a considerable weight percentage (~25%) of Ge along with the top electrode metal Pt (~19%); however, the X-ray peaks characteristic to Si and O are negligibly small. Further, the XRD plot, shown in Fig. 2(d), exhibits a single peak at $2\theta = 62.6^0$ confirming the formation of [400]-plane of Ge-nanowire [JCPDS 72-1089] on [400]-Si ($2\theta = 69.6^0$) substrate [JCPDS 01-0791], which also corroborates with the direction of crystal plane obtained from the SAED pattern.

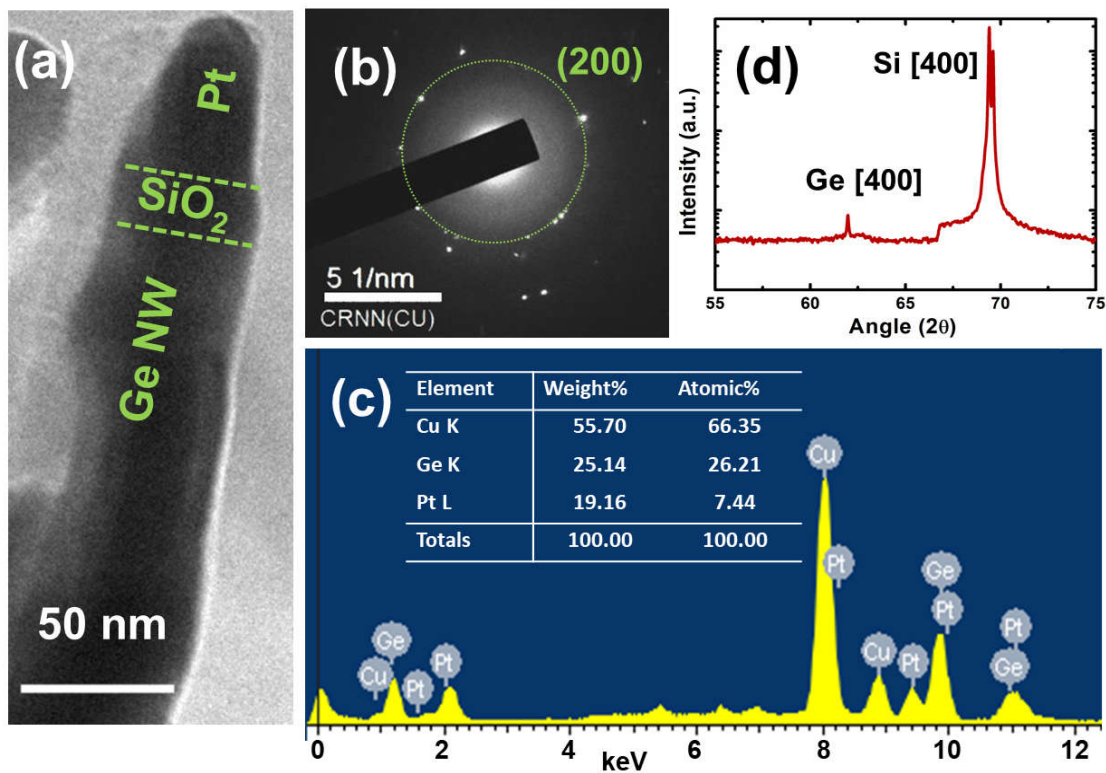

Fig. 3. (a) TEM image of the single nanowire MOS showing Ge (semiconductor), $SiO_2$ (oxide) and Pt (metal) regions where the nanowire radius and thickness of the oxide material are observed to be ~ 25 nm and ~20 nm respectively; (b) SAED pattern of the Ge-nanowire showing a ring with bright spots which corresponds to the [200]-plane of Ge; (c) Result of EDX-measurement performed in-situ under TEM along with the atomic and weight percentages of the relevant materials; (d) Plot of XRD profile of the patterned NWMOS confirming the Ge [400]-plane at 2θ = 62.6$^0$ on [400]-plane of Si substrate.

The capacitance-voltage (C-V) characteristics of a single Ge-NWMOS have experimentally been obtained by performing the measurement inside the FESEM chamber under high vacuum (~10$^{-6}$ mbars) at room temperature. The contacts for such measurement are made by tungsten (W) nano-probes attached to the micromanipulators connected to semiconductor characterization system (SCS) [23]. After careful setting of the probe contact, the electron beam is blanked during each run of measurement in order to avoid any perturbing contribution of charges from the background source of FESEM e-beam. However, it is worthy to mention that the SCS measuring apparatus generally overestimates the capacitance of ultra-scaled NWMOS significantly (in multiplicative order) due to large contact resistance (Si/Al/W). Such overestimated capacitance values obtained from the direct measurement can however be corrected to obtain the 'real capacitance' by utilizing the method of removing frequency dispersion associated with interface traps by reconstruction of lossy-MOS capacitance [24]. This methodology is utilized in the current work to estimate the C-V characteristics of single Ge-NWMOS [see SI] and plotted in Fig. 4(a) illustrating the high (1 MHz) and low (200 KHz) frequency responses of such device in the 'accumulation-to-inversion' region. It is worthy to mention that in the present case, the frequencies belonging to the order that corresponds to the inverse of minority carrier lifetime of Ge (~1 μs [25]) are considered to be 'low' and those above such range are assumed to be 'high'. It is apparent from Fig. 4(a) that the low frequency (200 kHz) C-V characteristics exhibit a 'step'-like behavior in the 'inversion' region (e.g., steps@ +2 V, +4 V) indicating strong 3D-confinement of electrons in the quantum well created at Ge-nanowire/SiO$_2$ interface at room temperature. The capacitance 'step's arise as a result of voltage driven lowering of the quantized states below Fermi level and to get occupied at room temperature discretely.

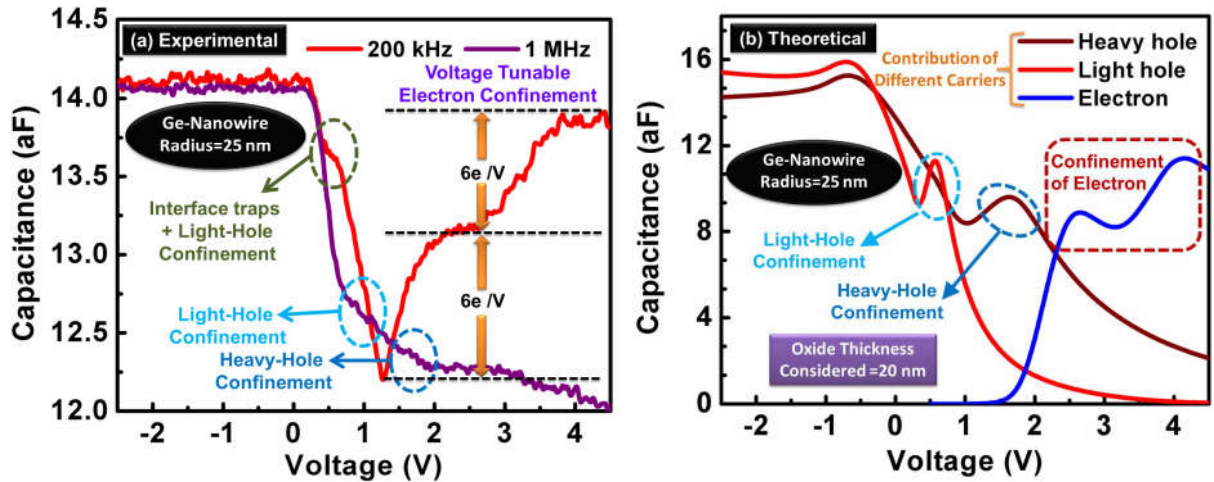

Fig. 4. (a) Plot of C-V characteristics measured in-situ under FESEM at room temperature for 200 kHz and 1 MHz frequencies showing accumulation-to-inversion region. The low frequency response of C-V curve exhibits step-like nature indicating 3D-confinement of electrons with ~6e per volt per step in the VTQD formation region. The humps created due to interface traps and hole (light/heavy) confinement are also shown; (b) Theoretical plot of C-V characteristics of a single nanowire MOS device of identical dimensions obtained from the analytical model based on NEGF formalism. In such plot the contributions of different carriers (heavy holes, light holes and electrons) in the net capacitance behavior are shown in corroboration with the experimental results.

Careful investigation shows that each of such major 'step's in the C-V curve exhibits a leap of ~6 electronic charge per volt and gets partially smoothened by a number of smaller convoluted 'step's. Such smaller 'steps' originate due to radial (i.e. geometrical) confinement whereas the major 'step's are attributed to axial (i.e. electrical) confinement tunable by the applied voltage. However, such 3D-quantization effect diminishes significantly at high frequency (exhibiting very small humps) since the probability of electrons (minority carriers) to respond within the time period larger than minority carrier lifetime is much lower [26]. On the other hand, C-V curves for both the low and high frequencies exhibit hump(s) while transiting from

'accumulation-to-inversion', the latter (i.e. high frequency humps) being reduced but not removed completely. Such hump arises from interface traps as well as the confinement of holes at Ge/p-Si interface (due to valence band off-set of 0.5 eV), the contribution of the former being diminished at high frequency, whereas that of the latter subsists [27, see SI]. Such confinement effect is contributed by both 'light' and 'heavy' holes leading to different humps convoluted into the C-V curve (Fig. 4(a)).

The nature of such variation of NWMOS capacitance with applied bias is investigated theoretically in the current work by solving quantum-electrostatic equations self-consistently based on non-equilibrium Green's function (NEGF) formalism [see SI]. The theoretical results are plotted in Fig. 4(b) illustrating the contributions of electron, heavy hole and light hole in the net capacitance of the device. Such theoretical model considers the Ge-nanowire to be 'active device' coupled to p-Si substrate as the 'reservoir'. The corresponding retarded Green's function of NWMOS for electrons in conduction band and holes in valence band are given by [see SI],

$$\underline{\underline{G}}^C(E_e) = \underset{\underline{\underline{\Sigma}}_{p-Si}(E_e) \to 0}{Lt} \left[ E_e \underline{\underline{I}} - \underline{\underline{E}}^C - \underline{\underline{\Sigma}}_{p-Si}(E_e) \right]^{-1} \qquad (1)$$

and

$$\underline{\underline{G}}^V(E_h) = \left[ E_h \underline{\underline{I}} - \underline{\underline{E}}^V - \underline{\underline{\Sigma}}_{p-Si}(E_h) \right]^{-1} \qquad (2)$$

where, $E_e$ and $E_h$ are the total electron and hole energy varying from band edges to +/- infinity, respectively. $\underline{\underline{E}}^C$ comprises of all the 3D-quantized energy states created in the voltage tunable quantum well in conduction band and $\underline{\underline{E}}^V$ represents the matrix consisting of 2D-confined valence sub-bands. The self-energy corresponding to coupling between the electrons and holes of

Ge-NW and p-Si substrate are represented by $\underline{\underline{\Sigma}}_{p-Si}(E_e)$ and $\underline{\underline{\Sigma}}_{p-Si}(E_h)$, respectively. It is worthy to mention that the electron self-energy is considerably small for two reasons: first, electrons are minority carriers in p-Si, and second, the quantized energy states within the well of nanowire exhibit significant mismatch with possible electron energy states in Si-substrate due to voltage induced band banding. Therefore, the density of states in conduction band, which is equivalent to the imaginary part of Green's function [28], results to be,

$$\underline{\underline{D}}^C(E_e) = 2\delta(E_e - \underline{\underline{E}}^C) \tag{3}$$

indicating the 3D-quantization of electrons. However, the occupancy of such electrons depends on whether the quantized states are near/below the p-Si Fermi level and controlled by the applied bias, which finally manifests as the 'step-like' behavior of device capacitance (Fig. 4 (a)-(b)). The comparison of such experimental and theoretical plots of capacitance suggests that the prominent hump at ~0.5 V for low frequency is originated due to the confinement of light holes, whereas the heavy hole confinement corresponds to the convoluted hump at ~1 V observed for high frequency. Such frequency response is attributed to the Ge-NW↔p-Si hole transmission lifetime [see SI],

$$\langle t \rangle \equiv \frac{\hbar}{2} \text{Im}[-\underline{\underline{\Sigma}}_{p-Si}(E_h)]^{-1} \tag{4}$$

which significantly depends on the mismatch of hole effective mass ($m_h*$) of Ge and Si.

The Ge-NW↔p-Si coupling is relatively stronger for 'heavy' holes in comparison to 'light' holes due to their lesser effective-mass-mismatch ($m_h^{HH}*(Ge) = 0.33$; $m_h^{LH}*(Ge) = 0.043$; $m_h^{HH}*(Si) = 0.49$; $m_h^{LH}*(Si) = 0.16$ [29]) resulting to larger broadening of hole sub-bands associated with reduced contact resistance [30], thereby exhibiting the corresponding frequency responses. To illustrate the overall charge quantization and the available energy space in present NWMOS device (for a given voltage of 3 V), the variation of local density of states (LDOS) with energy along nanowire axis from $SiO_2$/GeNW-interface is plotted for both the 3D-confined electrons and 2D-confined 'light' and 'heavy' holes in Fig 5(a) and 5(b), respectively.

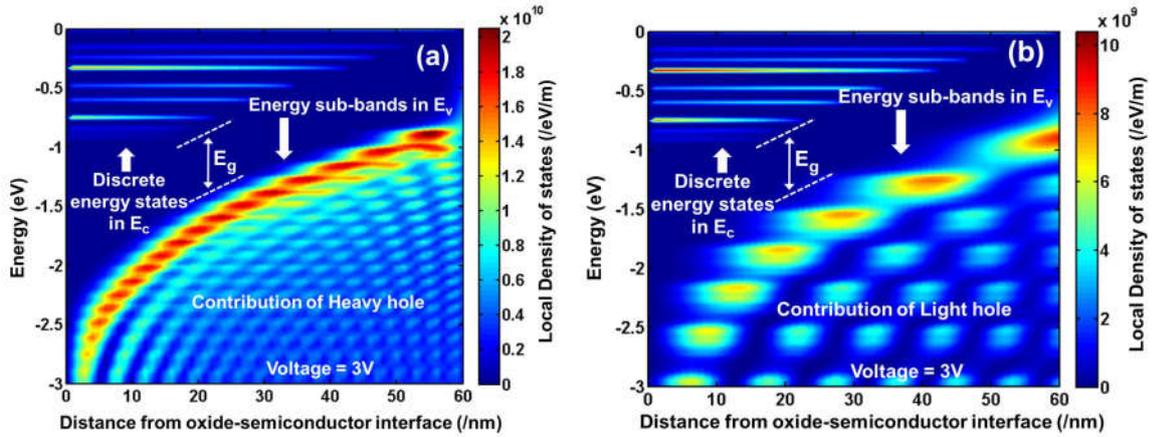

Fig. 5. Plots of local density of states in energy space with the distance from oxide/semiconductor interface for the 3D-confined electrons and 2D-confined holes: (a) heavy holes and (b) light holes. Applied voltage at the metal terminal is considered to be 3 V. The results are obtained from the NEGF based analytical model developed in the current work. The LDOS for electrons show spikes for created 3D-confined discrete states in the voltage tunable quantum well region whereas such LDOS for holes are broadened due to their free longitudinal motion towards the substrate. Such broadening is observed to depend significantly on the hole effective mass (i.e. light/heavy hole) mismatch between Si/Ge.

In conclusion, an array of voltage tunable quantum dots (VTQDs) based on ultra-scaled (radius ~25 nm) Ge-NWMOS devices with controlled variation of inter-spacing are fabricated, which exhibit strong quantum confinement at room temperature. The step-like behavior of capacitance shows ~6 electrons per step to be confined per volt in such device. The confinement properties of different types of carriers are theoretically investigated on the basis of NEGF formalism which provides physical insights about the quantum transport in such devices. The developed array of VTQDs can be utilized in advanced quantum technologies including novel photodevices as well as q-bit generation for quantum information processing.

**Acknowledgement**

Mr. Subhrajit Sikdar thanks the University Grant Commission (UGC), Government of India, for funding the fellowship through University of Calcutta. Dr. Basudev Nag Chowdhury would like to acknowledge Center of Excellence (COE), TEQIP Phase-II, World Bank for funding the postdoctoral fellowship. The authors also thank the Center of Excellence (COE) for the Systems Biology and Biomedical Engineering, and Center for Research in Nanoscience and Nanotechnology (CRNN), University of Calcutta, for providing the necessary infrastructural support. The authors also acknowledge the infrastructural support obtained from the WBDITE, West Bengal sponsored project.